\documentclass[twocolumn,showpacs,prl,amsmath,amssymb]{revtex4}
\RequirePackage[pdftex,pdfpagemode=none, pdftoolbar=true,
                pdffitwindow=true,pdfcenterwindow=true]{hyperref}
\usepackage[pdftex]{graphicx}
\usepackage{dcolumn}
\usepackage{bm}

\begin{document}
\title{Phonon renormalisation in doped bilayer graphene}
\author{A. Das$^1$}
\author{B. Chakraborty$^{1}$}
\author{S. Piscanec$^2$}
\author{S. Pisana$^2$}
\author{A. K. Sood$^{1}$}
\email{asood@physics.iisc.ernet.in}
\author{A. C. Ferrari$^2$}
\email{acf26@eng.cam.ac.uk}

\affiliation{$^{1}$Department of Physics, Indian Institute of Science, Bangalore 560012,India\\
 $^{2}$Engineering Department, Cambridge University, Cambridge CB3 0FA, UK}


\begin{abstract}
We report phonon renormalisation in bilayer graphene as a function of doping. The Raman
G peak stiffens and sharpens for both electron and hole doping, as a result of the
non-adiabatic Kohn anomaly at the $\Gamma$ point. The bilayer has two conduction and valence
subbands, with splitting dependent on the interlayer coupling. This
results in a change of slope in the variation of G peak position with doping, which allows a direct
measurement of the interlayer coupling strength.
\end{abstract}
\maketitle

Graphene is the latest carbon allotrope to be
discovered~\cite{NovScience2004,NovNature2005,ZhangNature2005,revchar,GeimRev}. Near-ballistic
transport at room temperature and high carrier mobilities\cite{NovNature2005,ZhangNature2005, MorozovNov(2007),revchar,GeimRev,andrei,kimmob},
make it a potential material for nanoelectronics~\cite{Lemme,kimribbon, avouris}, especially for
high frequency applications. It is now possible to produce areas exceeding thousands of square
microns by means of micro-mechanical cleavage of graphite. An ongoing effort is being devoted to
large scale deposition and growth on different substrates of choice.

Unlike single layer graphene (SLG), where electrons disperse linearly as massless Dirac
fermions\cite{NovScience2004,NovNature2005,ZhangNature2005,revchar,GeimRev}, bilayer graphene (BLG)
has two conduction and valence bands, separated by $\gamma$$_{1}$, the interlayer
coupling\cite{mccann1,andobi}. This was measured to be$\sim$0.39eV by angle resolved photoelectron
spectroscopy\cite{arpes}. A gap between valence and conduction bands could be opened and tuned by
an external electric field ($\sim$100meV for$\sim$10$^{13}$cm$^{-2}$doping)\cite{McCann,Neto
Biased}, making BLG a tunable-gap semiconductor.
\begin{figure}
\includegraphics[width=80mm]{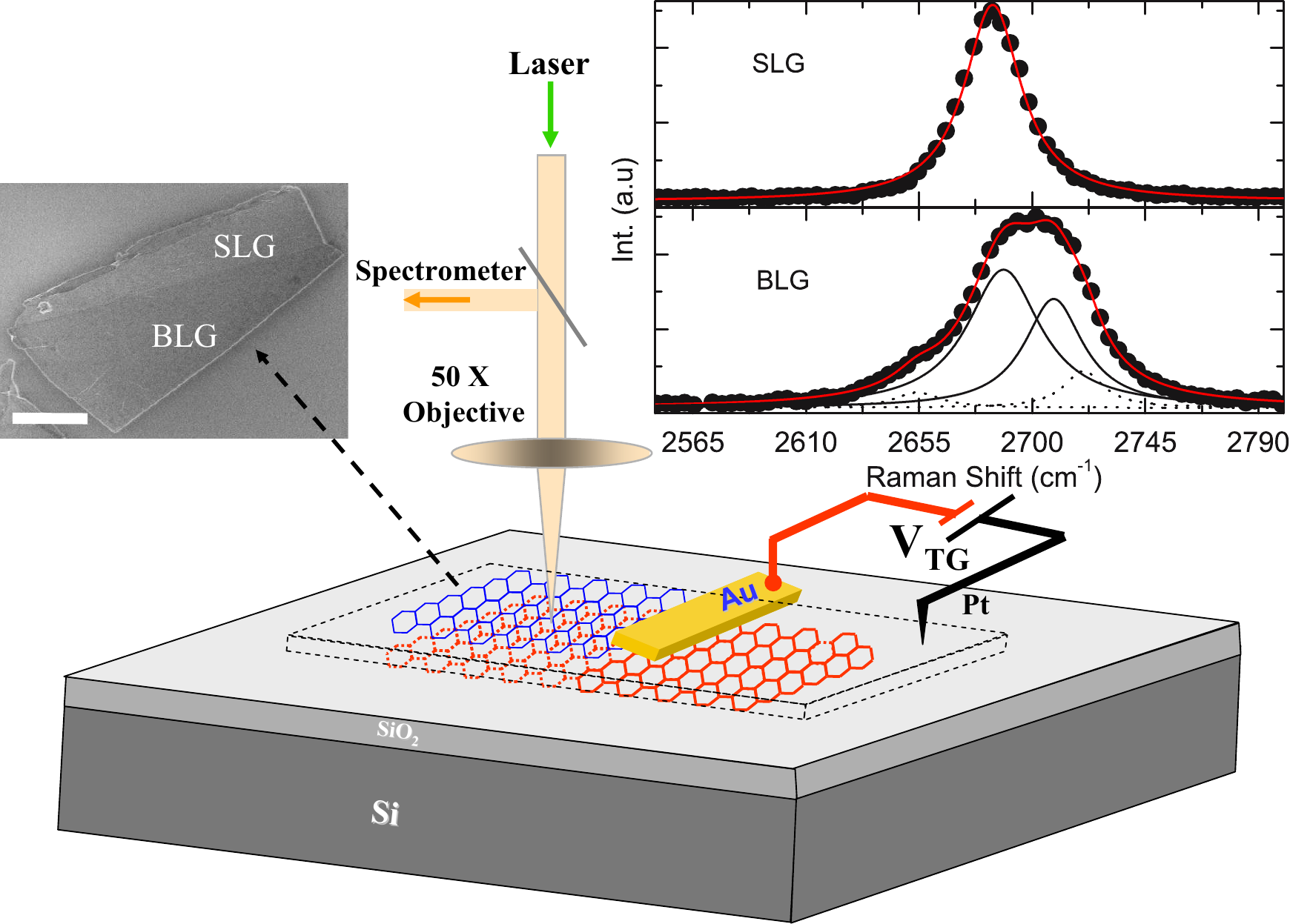}
\caption{(color online). Experimental setup. The black dotted box on SiO$_{2}$ indicates the polymer electrolyte (PEO + LiClO$_{4}$). The left
inset shows an SEM image of the SLG and BLG. Scale bar: 4$\mu$m. The right inset, the 2D Raman band}
\label{Figure1}
\end{figure}

Graphene can be identified in terms of number and orientation of layers by means of elastic and
inelastic light scattering, such as Raman\cite{FerrariPrl2006} and Rayleigh
spectroscopies\cite{CasiraghiNL,GeimAPL}.  Raman spectroscopy also allows monitoring of doping and
defects\cite{SSCReview,dasgra,SimoneNaturematerials2007,YanPrl2007,CasAPL,revchar,dasBMS}. Indeed, Raman
spectroscopy is a fast and non-destructive characterization method for carbons\cite{acftrans}. They
show common features in the 800-2000 cm$^{-1}$ region: the G and D peaks, around 1580 and 1350
cm$^{-1}$, respectively. The G peak corresponds to the $E_{2g}$ phonon at the Brillouin zone center
($\Gamma$). The D peak is due to the breathing modes of sp$^2$ atoms and requires a defect for its
activation\cite{tuinstra,Ferrari00,ThomsenPrl2000}. The most prominent feature in SLG is the second
order of the D peak: the 2D peak\cite{FerrariPrl2006}. This lies at $\sim$ 2700 cm$^{-1}$ and
involves phonons at \textbf{K}+$\Delta$\textbf{q}\cite{FerrariPrl2006,SSCReview}.
$\Delta$\textbf{q} depends on the excitation energy, due to double-resonance, and the linear
dispersion of the phonons around \textbf{K}\cite{ThomsenPrl2000,Piscanec2004,FerrariPrl2006}. 2D is
a single peak in SLG, whereas it splits in four in BLG, reflecting the evolution of the band
structure\cite{FerrariPrl2006}. The 2D peak is always seen, even when no D peak is present, since
no defects are required for overtone activation.

In SLG, the effects of back and top gating on G-peak position (Pos(G)) and Full Width at Half
Maximum (FWHM(G)) were reported in Refs\cite{SimoneNaturematerials2007,YanPrl2007,dasgra}. Pos(G)
increases and FWHM(G) decreases for both electron and hole doping. The G peak stiffening is due to
the non-adiabatic removal of the Kohn-anomaly at
$\Gamma$\cite{Lazzeri2006,SimoneNaturematerials2007}. FWHM(G) sharpening is due to blockage of
phonon decay into electron-hole pairs due to the Pauli exclusion principle, when the electron-hole
gap is higher than the phonon energy\cite{LazPRB2006,SimoneNaturematerials2007}, and saturates for
a Fermi shift bigger than half phonon energy\cite{SimoneNaturematerials2007,YanPrl2007,LazPRB2006}.
A similar behavior is observed for the LO-G$^{-}$ peak in metallic nanotubes\cite{dasnt}, for the
same reasons. The conceptually different BLG band structure is expected to renormalize the phonon
response to doping differently from SLG\cite{andosi,andobi}. Here we prove this, by investigating
the effect of doping on the BLG G and 2D peaks. The G peak of doped BLG was recently
investigated\cite{kimBi}, and reproduced that of SLG, due to the very low doping
range($\sim5\times10^{12}cm^{-2}$), not enough to cross the second BLG subband. Here we reach much
higher values ($\sim5\times$$10^{13}$cm$^{-2}$), probing the further renormalisation resulting from
crossing to the second BLG subband.

\begin{figure}
\includegraphics[width=80mm]{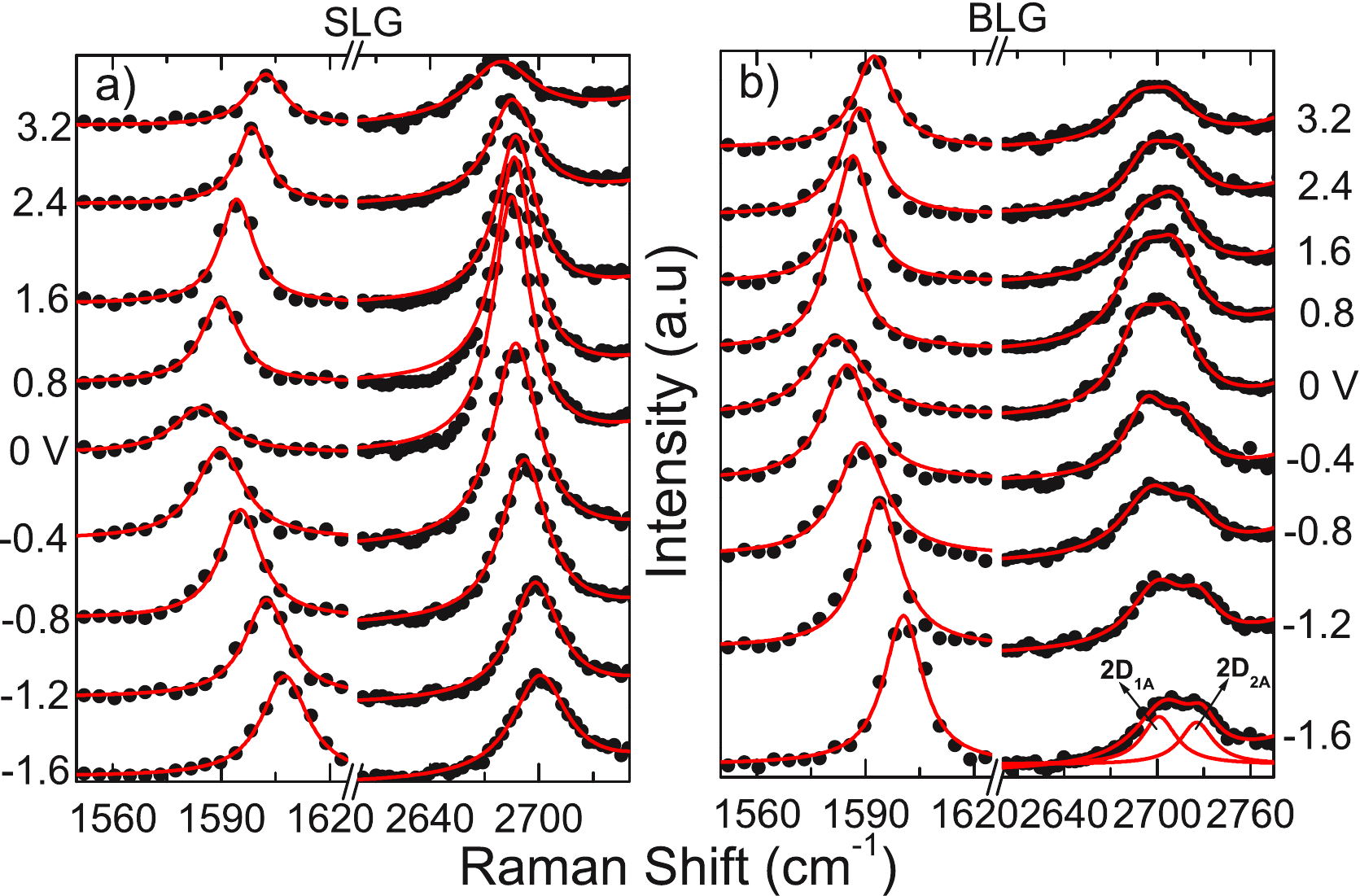}
\caption{(color online). Raman spectra of (a) SLG; (b) BLG at several
V$_{TG}$. Red lines fits to the experimental data.}
\label{Figure2}
\end{figure}

We recently demonstrated a SLG top-gated by polymer electrolyte\cite{dasgra}, able to span a large
doping range, up to$\sim$5$\times$10$^{13}$cm$^{-2}$\cite{dasgra}. This is possible because the
nanometer thick Debye layer\cite{shim,liu,dasgra} gives a much higher gate capacitance compared to
the usual 300nm SiO$_{2}$ back gate\cite{GeimRev}. We apply here this approach to BLG.
Fig.\ref{Figure1} shows the scheme of our experiment. A sample is produced by micromechanical
cleavage of graphite. This consists of a SLG extending to a BLG, as proven by the
characteristic SLG and BLG 2D peaks in the inset of Fig.\ref{Figure1}\cite{FerrariPrl2006}. An Au
electrode is then deposited by photolithography covering both SLG and BLG, Fig.\ref{Figure1}. Top
gating is achieved by using a solid polymer electrolyte consisting of LiClO$_{4}$ and polyethelyne
oxide (PEO) in the ratio 0.12:1\cite{dasgra}. The gate voltage is applied by placing a platinum
electrode in the polymer layer. Note that the particular shape of our sample, consisting of a BLG, with a protruding SLG, ensures the top gate to be effectively applied to
both layers at the same time. This would not necessarily be the case for a monolithic BLG,
where, due to screening effects, the gate would give a separate evolution of the Raman spectra of
the top and bottom layers\cite{dasum}. Measurements are done with a WITEC confocal (X50
objective) spectrometer with 600 lines/mm grating, 514.5 nm excitation, at$<$1mW to avoid heating.
For a given top gate voltage, V$_{TG}$, spectra are recorded after 10 mins. Figs.\ref{Figure2}(a,b)
plot the spectra as a function of V$_{TG}$. We use Voigt functions to fit the G peak in both SLG
and BLG. The SLG 2D band is fitted to one Lorentzian. The BLG 2D band is fitted to four
Lorentzians,2D$_{1A}$,2D$_{1B}$,2D$_{2A}$,2D$_{2B}$~\cite{FerrariPrl2006}, Fig.\ref{Figure1}. As
previously discussed, two of these, 2D$_{1A}$ and 2D$_{2A}$, are much
stronger\cite{FerrariPrl2006}. Thus, we focus on these.

To get a quantitative understanding, it is necessary to convert V$_{TG}$ into a E$_F$ shift. For
electrolytic gating, the chemical potential is
eV${_{TG}}=E^{SLG}_{F}+e\phi{^{SLG}}=E^{BLG}_{F}+e\phi{^{BLG}}$. The electrostatic potential
$\phi=\frac{ne}{C{_{TG}}}$ is determined by the geometrical capacitance $C{_{TG}}$ and carrier
concentration $n$ ($e$ is the electron charge), while $E_{F}$/e by the chemical
(quantum) capacitance of graphene. For SLG, $n{^{SLG}} = \mu{E^{2}_{F}}$, where
$\mu=\frac{g{_{s}}g{_{v}}}{4\pi\gamma^{2}}=\frac{1}{\pi({\hbar}v{_{F}}){^{2}}}$, g$_{s}$=g$_{v}$=2
are spin and valley degeneracies, $\gamma=\frac{\sqrt{3}}{2}\gamma_0 a$, with $\gamma_0$ the
nearest-neighbor tight binding parameter, $a$ the graphene lattice parameter, and $v_{F}$ is
the Fermi velocity. Thus:
\begin{equation}
eV{_{TG}}=E{_{F}}+{\nu}E^{2}_{F} \label{Ef2VgSLG}
\end{equation}
For BLG\cite{andobi,note_andobi,Review Elec} $n^{BLG}$=$\mu[\gamma{_{1}}{E{_{F}}}+E^{2}_{F}]$ for
$E{_{F}}<\gamma{_{1}}$ and $n^{BLG}=2\mu{E^{2}_{F}}$ for $E{_{F}}>\gamma{_{1}}$. Thus:
\begin{eqnarray}\label{Ef2VgBLG}
eV{_{TG}}=&(1+\nu\gamma_1)E{_{F}}+\nu E^{2}_{F}, &        E{_{F}}<\gamma_1 \\
& = E{_{F}}+ 2\nu E^{2}_{F}, &        E{_{F}} >
\gamma{_{1}}\nonumber
\end{eqnarray}
where $\nu=\frac{e{^{2}}}{{\pi}C{_{TG}}({\hbar}v{_{F}}){^{2}}}$. We take
$C_{TG}=2.2\times10^{-6}$Fcm$^{-2}$\cite{dasnt}, and $\gamma$$_{1}$=0.39eV constant with doping
(since its variation for n up to$\sim$10$^{13}$cm$^{-2}$ is $<$5$\%$)\cite{arpes,McCann}).
Eqs.\ref{Ef2VgSLG},\ref{Ef2VgBLG} then give $E_F$ as a function of $V_{TG}$.

\begin{figure}
\includegraphics[width=80mm]{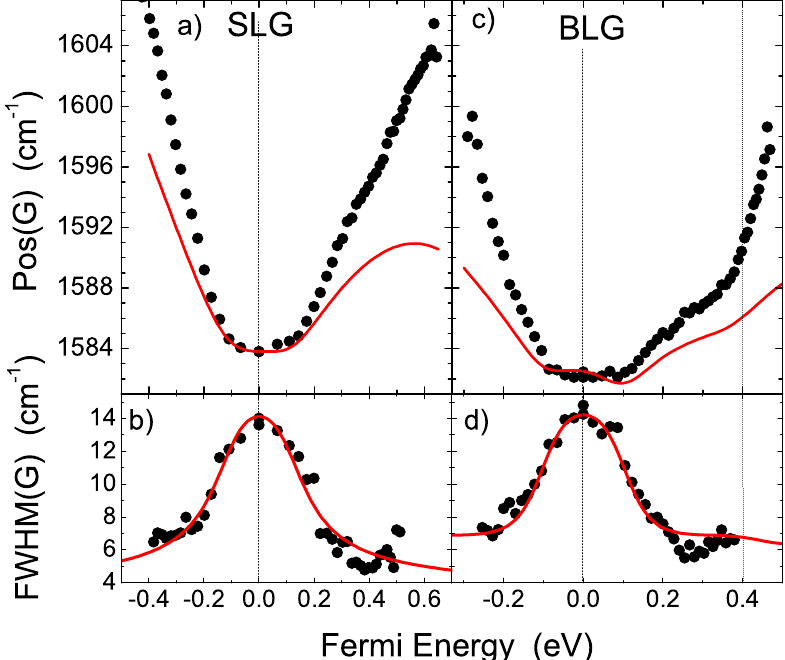}
\caption{(color online). Pos(G) for (a) SLG; (c) BLG as a
function of Fermi energy. FWHM(G) of (b) SLG;(d) BLG as a
function of Fermi energy. Solid lines:theoretical
predictions.} \label{Figure3}
\end{figure}

Fig.\ref{Figure3} plots the resulting Pos(G), FWHM(G) as a function of $E_F$. In SLG, Pos(G) does not increase up to $E_{F}\sim$0.1eV ($\sim\hbar\omega_0$/2), where $\omega_0$ is the frequency of
the $E_{2g}$ phonon in the undoped case ($\hbar\omega_0/(2\pi\hbar
c)$=Pos(G$_{0}$), with $c$ the speed of light), and
then increases with $E_F$. Fig.\ref{Figure3}b,d indicate that in SLG and BLG, FWHM(G) decreases for
both electron and hole doping, as expected since phonons decay into real electron-hole pairs when
$E_{F}<\hbar\omega_0$/2\cite{SimoneNaturematerials2007}. Fig.\ref{Figure3}c plots Pos(G) of
BLG.(i) Pos(G) does not increase until $E_{F}\sim$0.1eV ($\sim\hbar\omega_0$/2).(ii) Between 0.1 and 0.4eV, the BLG slope R=$\frac{dPos(G)}{dE{_{F}}}$ is smaller than the SLG one.(iii) A kink is observed in Fig.3b
at E$_{F}\sim$0.4eV.(iv)Beyond E$_{F}>$0.4eV the slope is larger than in SLG.(v) The kink position does not significantly depend on $\gamma_1$ used to convert $V_{TG}$ in $E_F$(e.g. a $\sim66\%$ change in $\gamma_1$ modifies E$_F$ by $\sim6\%$).

These trends can be explained by considering the effects doping on the phonons:(i) a change of the equilibrium lattice parameter with a consequent ``static'' stiffening/softening, $\Delta Pos(G)^{st}$;(ii) the onset of ``dynamic'' effects beyond the adiabatic
Born-Oppenheimer approximation, that modify the phonon dispersion close to the Kohn anomalies, $\Delta Pos(G)^{dyn}$\cite{SimoneNaturematerials2007,Lazzeri2006}. Thus, the total phonon
renormalization can be written as\cite{SimoneNaturematerials2007,Lazzeri2006}:
\begin{equation}
{\rm Pos}(G_{E_F})-{\rm
Pos}(G_0)=\Delta Pos(G)=\Delta Pos(G)^{st}+\Delta Pos(G)^{dyn}
\label{DeltaPos}
\end{equation}
For SLG, we compute $\Delta Pos(G)^{st}$ by converting $E_F$
into the corresponding electron density $n^{SLG}$, then using Eq.3 of Ref.\cite{Lazzeri2006}.
For BLG, we assume $n^{BLG}$ equally distributed on the two
layers, each behaving as a SLG with an electron concentration $n^{BLG}/2$. Eq.3 of Ref.\cite{Lazzeri2006} is then used to compute
$\Delta Pos(G)^{st}$ for BLG. $\Delta Pos(G)^{dyn}$ is calculated from the phonon self-energy $\Pi$\cite{Pickett}:
\begin{equation}
\hbar\Delta Pos(G)^{dyn}={\rm Re}[\Pi(E_{F})-\Pi(E_{F}=0)].
\label{DeltaOmegaDyn}
\end{equation}
The electron-phonon coupling (EPC) contribution to FWHM(G) is given by\cite{Pickett,Allen,Note_FWHM}:
\begin{equation}
{\rm FWHM(G)}^{EPC}=2 {\rm Im}[\Pi(E_F)] \label{EqFWHM}
\end{equation}
The self-energy for the $E_{2g}$ mode at
$\bm \Gamma$ in SLG is\cite{SimoneNaturematerials2007,Lazzeri2006}:
\begin{equation}
\Pi(E_F)^{SLG}=\alpha'\int^{\infty}_{-\infty}\frac{f(\epsilon)-f(-\epsilon)}{2\epsilon+\hbar\omega_{0}+i\delta}|\epsilon|d\epsilon,
\label{SelfEnergySLG}
\end{equation}
while for BLG it is given by~\cite{andobi}:
\begin{eqnarray}
\Pi(E_F)^{BLG} &=& \alpha'\int^{\infty}_{0}{\gamma{^{2}}}kdk{\sum_{s,s{^{\prime}}}}{\sum_{j,j{^{\prime}}}}{\phi^+{_{jj{^{\prime}}}}} \nonumber \\
&{\times}&
\frac{[f(\epsilon_{sjk})-f(\epsilon_{s^{\prime}j^{\prime}k})][\epsilon_{sjk}-\epsilon_{s^{\prime}j^{\prime}k}]}{(\epsilon_{sjk}-\epsilon_{s^{\prime}j^{\prime}k})^{2}-(\hbar\omega_0+i\delta)^{2}}
\label{bi}
\end{eqnarray}
where $\alpha'=\frac{\hbar A_{uc} EPC(\Gamma)^2}{\pi M \omega_0
(\hbar v_F)^2}$, $A_{uc}=5.24$~\AA$^2$ is the graphene unit-cell
area, $M$ is the carbon atom mass, $f(\epsilon)=1/[\exp(\frac{\epsilon-E_F}{k_BT})+1]$ is the
Fermi-Dirac distribution, $\delta$ is a broadening factor
accounting for charge inhomogeneity, EPC($\Gamma$) is the electron
phonon coupling\cite{note01}. $s=\pm$1 and $s^{\prime}$=$\pm$1
label the conduction (+1) and valence (-1) bands, while $j=1,2$
and $j^{\prime}$=1,2 label the two parabolic subbands.
$\epsilon_{sjk}$ is computed from Eq.2.8 of Ref.\cite{andobi}, and $\phi^+_{jj'}$ is given by Eq.3.1 of Ref.\cite{andobi}. By using Eqs.\ref{SelfEnergySLG},\ref{bi} in
Eqs.\ref{DeltaOmegaDyn},\ref{EqFWHM}, we get $\Delta Pos(G)^{dyn}$,FWHM(G)$^{EPC}$ for SLG and BLG.

To compare Eqs.\ref{DeltaPos},\ref{EqFWHM} with the
experimental data, we use $\alpha'=4.4\times10^{-3}$(obtained
from the DFT values of EPC($\Gamma$) and $v_F$\cite{Piscanec2004,SimoneNaturematerials2007}), the experimental $\hbar\omega_0$ for SLG and BLG, and T=300K. $\delta$ is fitted from the experimental
FWHM(G) to FWHM(G)=FWHM(G)$^{EPC}$+FWHM(G)$^{0}$, with FWHM(G)$^{0}$ a constant accounting for non-EPC effects (e.g. resolution and anharmonicity). For SLG (BLG) we get $\delta=0.13$eV (0.03eV) and FWHM(G)$^0$=4.3cm$^{-1}$ (5.1cm$^{-1}$). These $\delta$ values are then used to compute Pos(G). Note that the relation between n and E$_{F}$ implies that charge inhomogeneity causes different E$_F$ broadening in SLG and BLG (e.g. $\delta$n$\sim$10$^{12}$cm$^{-2}$ would give 0.13eV and 0.03eV in SLG and BLG,respectively).
\begin{figure}
\includegraphics[width=65mm]{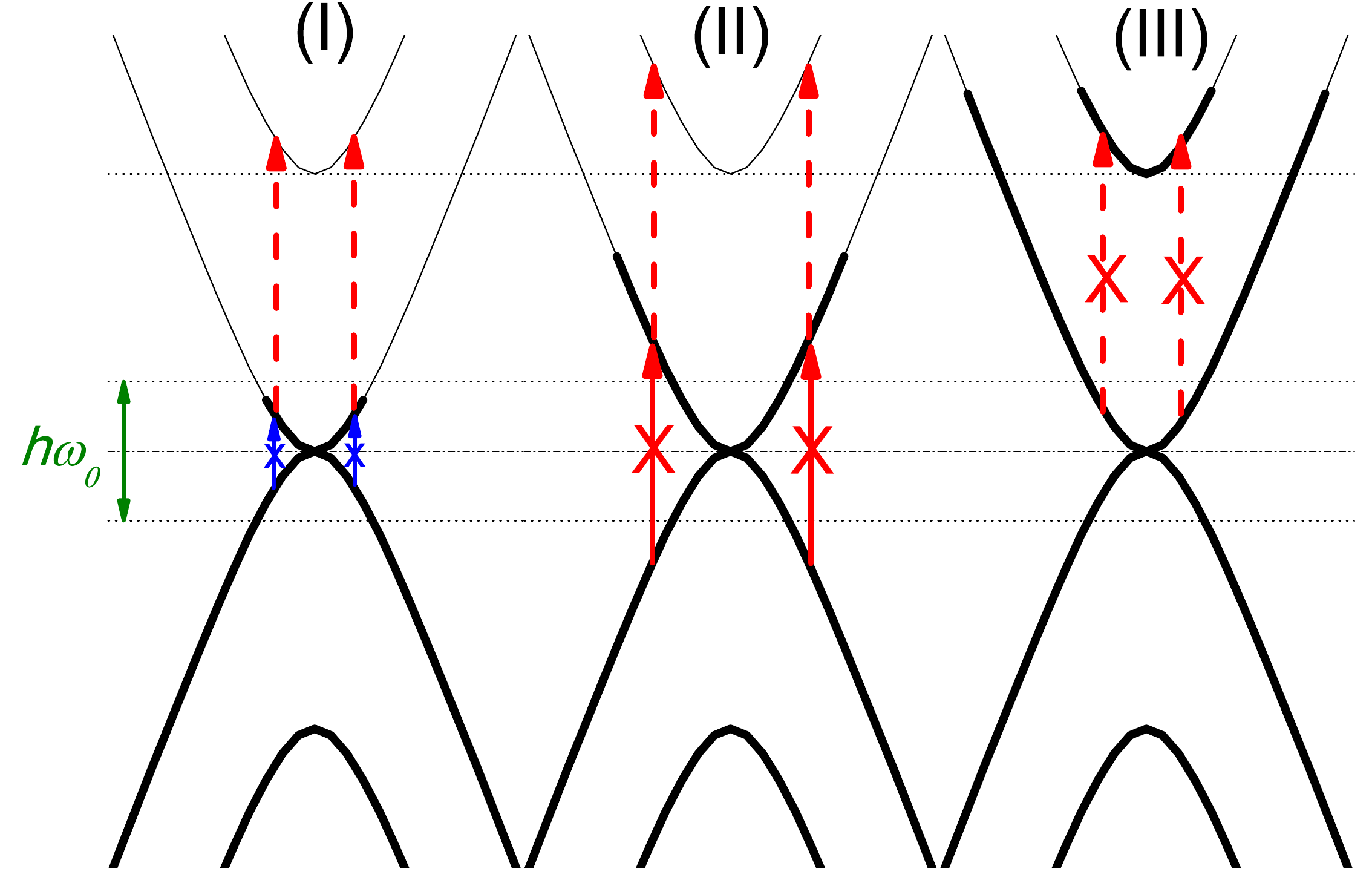}
\caption{(color online). Phonon renormalization for BLG: (I) $E_F<\hbar\omega_0$, (II) $\hbar\omega_0<E_F<\gamma_1$, (III) $E_F>\gamma_1$. Blue and red arrows correspond respectively to positive and negative contributions to $\Pi$. Solid and dashed arrows correspond to interband and intraband processes respectively.}
\label{Figure4}
\end{figure}

The solid lines in Fig.~\ref{Figure3} are the theoretical Pos(G) and FWHM(G) trends. The experimental and theoretical FWHM(G) are in excellent agreement, as expected since the latter was fitted to the former. The theoretical Pos(G) captures the main experimental features. In particular, the flat dependence for $|E_f|<0.1$ eV in both SLG and BLG, and the kink at $\sim0.4$ eV in BLG. This kink is the most striking difference between SLG and BLG. It is the signature of
the second subband filling in BLG. Indeed, a shift of E$_F$, by acting on $f(\epsilon)$ in Eq.\ref{bi}, modifies the type and number of transitions contributing to $\Pi$. The only transitions giving a positive contribution to $\Pi$ are those for which $|\epsilon_{s,j,k}-\epsilon_{s',j',k}|<\hbar\omega_0$, i.e. a subset of those between $(s=-1;j=1)$ and $(s=1;j=1)$  (interband transitions, solid blue lines in Fig.~\ref{Figure4}). Interband transitions with $|\epsilon_{s,j,k}-\epsilon_{s',j',k}|>\hbar\omega_0$ (solid red lines in Fig.~\ref{Figure4}) and all intraband (between $(s=\pm 1;j=1)$ and $(s=\pm 1,j=2)$, dashed red lines in Fig.~\ref{Figure4}) contribute to $\Pi$ as negative terms. It is convenient to distinguish three different cases: (I), $|E_F|<\hbar\omega_0$, (II) $\hbar\omega_0<|E_F|<\gamma_1$, and (III) $|E_F|>\gamma_1$. For simplicity let us assume $E_F>0$ (the same applies for $E_F<0$).In case (I), positive contributions from interband transitions are suppressed, and new negative intraband transitions are created. This results in strong phonon softening at low temperatures\cite{kimBi}. At T=300K, these effects are blurred by the fractionary occupation of the electronic states, resulting in an almost doping independent phonon energy (see Fig.\ref{Figure3}b). In case (II), a shift of $E_F$ suppresses negative interband contributions and creates new negative intraband transitions. By counting their number and relative weight (given by $\Phi_{jj'}/(\epsilon_{s,j,k}-\epsilon_{s',j',k})$), one can show that interband transitions outweight intraband ones, resulting in phonon hardening. Case (III) is similar to (II), with the difference that the second subband filling suppresses negative intraband transitions at \textbf{k}$\sim$\textbf{K}, further enhancing the phonon hardening. It is also possible to demonstrate that, for T and $\delta \to0$, the slope of $\Delta Pos(G)^{dyn}$ just above $E_F=\gamma_1$ is double than that just below. Thus, the kink in Fig.3 is a direct measurement of the interlayer coupling strength from Raman spectroscopy.

In SLG the intensity ratio of 2D and G, I(2D)/I(G), has a strong
dependence on doping\cite{dasgra}. Fig.5a plots I(2D)/I(G) as a function of doping. For BLG we take the highest amongst 2D$_{1A}$ and 2D$_{2A}$. The SLG dependence reproduces our previous results\cite{dasgra}.
However, we find an almost constant ratio in BLG. Fig.5b plots the doping dependence of Pos(2D) in SLG, and Pos(2D$_{1A}$), Pos(2D$_{2A})$ in BLG. To a first approximation, this is governed by lattice relaxation, which explains the overall stiffening for hole doping and softening for electron
doping\cite{dasgra}. A quantitative understanding is yet to emerge, and beyond DFT many body effects need be considered.
\begin{figure}
\includegraphics[width=68mm]{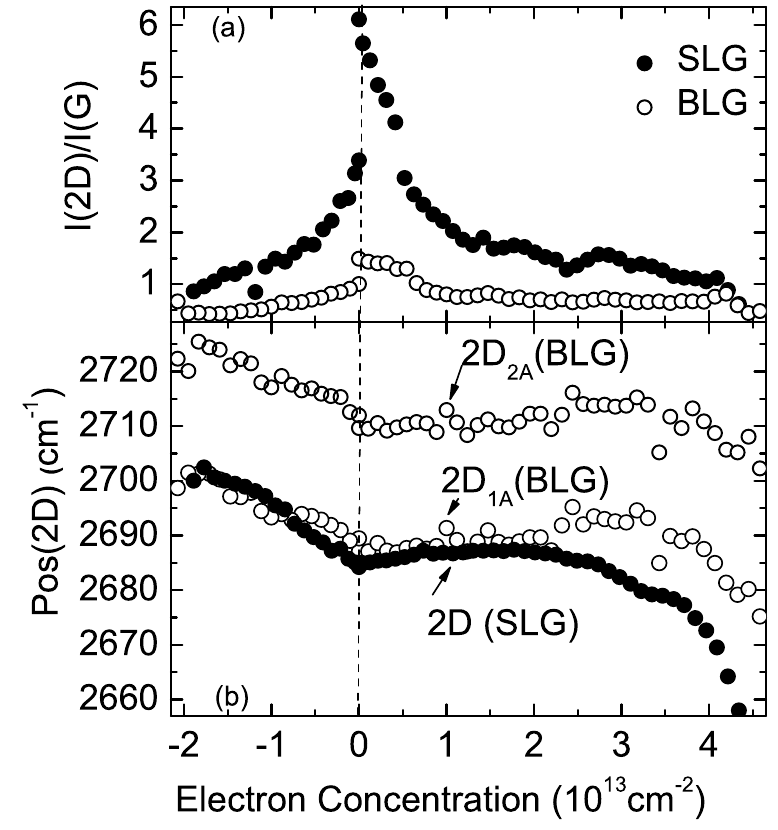}
\caption{(a) Ratio of 2D and G peaks intensities for SLG (solid circles) and BLG (open circles) as a function of n. (b) Position of 2D for SLG (solid circles) and 2D main components for BLG (open circles) as a function of n.}
\label{Figure5}
\end{figure}

To conclude, we have simultaneously measured the behavior of optical phonons in single and bilayer
graphene as a function of doping. In the latter, the G peak renormalizes as the Fermi energy moves from the 1st to the 2nd subband, allowing a direct measurement of $\gamma_{1}\sim$0.4eV.

We thank D. Basko for useful and stimulating discussions. AKS acknowledges funding from the Department of Science and Technology, India, SP from Pembroke College and the Maudslay society, ACF from The Royal Society and The
Leverhulme Trust.


\begin{thebibliography}{10}

\bibitem{NovScience2004} K.S. Novoselov et al. Science \textbf{306}, 666 (2004)

\bibitem{NovNature2005}K.S. Novoselov, et al. Nature \textbf{438},197 (2005).

\bibitem{ZhangNature2005} Y. Zhang et al. Nature, \textbf{438}, 201 (2005).

\bibitem{revchar} J. C. Charlier et al. Topics Appl. Phys. \textbf{111}, 673 (2008)

\bibitem{GeimRev}A. K. Geim, K. S. Novoselov; Nat. Mater., \textbf{6}, 183 (2007).

\bibitem{MorozovNov(2007)} S. V. Morozov et al. Phys. Rev. Lett., \textbf{100} 016602 (2008).

\bibitem{andrei} X. Du et al. cond mat. arXiv:0802.2933

\bibitem{kimmob}  K. I. Bolotin et al arXiv:0802.2389; arXiv:0805.1830

\bibitem{Lemme} M. C. Lemme et al. IEEE El. Dev. Lett. \textbf{28}, 282 (2007).

\bibitem{kimribbon} M. Y. Han et al. Phys. Rev. Lett. \textbf{98}, 206805 (2007).

\bibitem{avouris} Z. Chen et al. Physica E, \textbf{40}, 228 (2007).

\bibitem{mccann1} E.McCann,V.I.Falko, Phys. Rev. Lett.,\textbf{96} 086805 (2006)

\bibitem{andobi} T. Ando, J. Phys. Soc. Jpn. \textbf{76}, 104711 (2007).

\bibitem{arpes} T. Ohta et al. Science \textbf{313}, 951 (2007).

\bibitem{McCann} E. McCann, Phys. Rev. B \textbf{74}, 161403 (R) (2006).

\bibitem{Neto Biased}  E. V. Castro et al. Phys. Rev. Lett. \textbf{99}, 216802 (2007).

\bibitem{FerrariPrl2006} A. C. Ferrari et al. Phys. Rev. Lett. \textbf{97}, 187401 (2006).

\bibitem{CasiraghiNL} C. Casiraghi et al. Nano. Lett. \textbf{7}, 2711 (2007).

\bibitem {GeimAPL} P. Blake et al. Appl. Phys. Lett. \textbf{91}, 063124 (2007).

\bibitem{SimoneNaturematerials2007} S. Pisana et al., Nature Mat. \textbf{6},198 (2007).

\bibitem{YanPrl2007} J. Yan et al. Phys.Rev.Lett. \textbf{98}, 166802 (2007).

\bibitem{CasAPL} C. Casiraghi et al. Appl. Phys. Lett. \textbf{91}, 233108 (2007)

\bibitem{SSCReview} A. C. Ferrari, Solid State Comm. \textbf{143}, 47 (2007).

\bibitem{dasgra} A. Das et al., Nature Nanotech. \textbf{3}, 210 (2008).

\bibitem{dasBMS} A. Das et al., cond-mat/0710.4160 (2007).

\bibitem{acftrans} A. C. Ferrari, J. Robertson (eds),
Raman spectroscopy in carbons: from nanotubes to diamond, Theme Issue, Phil. Trans. Roy. Soc. A \textbf{362}, 2267-2565 (2004).

\bibitem{tuinstra}
F.Tuinstra,J.L. Koenig, J. Chem. Phys.\textbf{53}, 1126 (1970).

\bibitem{Ferrari00}
A.C. Ferrari, J. Robertson Phys. Rev. B \textbf{61}, 14095 (2000);
{\it ibid.} 64, 075414 (2001).

\bibitem{ThomsenPrl2000} C. Thomsen, S. Reich, Phys. Rev.Lett. \textbf{85}, 5214 (2000).

\bibitem{Piscanec2004} S. Piscanec et al. Phys. Rev. Lett. \textbf{93}, 185503 (2004).

\bibitem{Lazzeri2006} M. Lazzeri, F. Mauri, Phys. Rev. Lett. \textbf{97}, 266407 (2006).

\bibitem{LazPRB2006} M. Lazzeri et al. Phys. Rev. B \textbf{73}, 155426 (2006).

\bibitem{dasnt} A. Das et al. Phys Rev Lett. \textbf{99}, 136803 (2007).

\bibitem{andosi} T. Ando, J. Phys. Soc. Jpn. \textbf{75}, 124701 (2006).


\bibitem{kimBi}  J. Yan et al. cond-mat/0712.3879v1 (2007).


\bibitem{shim} K. T. Nguyen, Phys. Rev. Lett. \textbf{98}, 145504 (2007).

\bibitem{liu} C. Lu et al. Nano Lett. \textbf{4}, 623 (2004).

\bibitem{dasum} A. Das, A. C. Ferrari, A. K. Sood, unpublished (2008)

\bibitem{note_andobi} Note that in Fig.3 of Ref.\cite{andobi} both SLG density
    of states and electron concentration are multiplied by a factor 2.

\bibitem{Review Elec} A. H. Castro Neto et al. cond-mat/0709.1163v1 (2007).


\bibitem{Pickett}W.E.Pickett,P.B.Allen, Phys. Rev. B\textbf{16}, 3127 (1977).

\bibitem{Allen} P. B. Allen, Phys. Rev. B \textbf{6}, 2577
(1972).

\bibitem{Note_FWHM} Note that the phonon self-energy imaginary part corresponds to G half
width at half maximum, HWHM(G), as for Eq.8 in Ref.\cite{Allen}. Thus, the factor 2 to compute FWHM(G) in Eq.5. This is sometimes neglected in literature. For example, $\Delta\Gamma$ in Eq.1 of Ref.\cite{YanPrl2007} represents HWHM(G), and not FWHM(G). Ref.\cite{YanPrl2007} then compares this with FWHM(G) calculated in Eq.3 of Ref.\cite{LazPRB2006}, finding $D^2/4=\langle D^2_{\bf \Gamma}\rangle_F$. However, the correct relation should be $D^2/2=\langle D^2_{\bf \Gamma}\rangle_F$. Because of this, the coupling constant of Ref.\cite{YanPrl2007} is $\lambda=2\alpha'$ instead
of $\lambda=\alpha'$. Similarly, "broadening" in Figs.4,6 of Ref.\cite{andobi} and Fig.4 of Ref.\cite{andosi} is HWHM(G), not FWHM(G). Also, Fig.6 in Ref.\cite{andoSWNTs} mistakenly compares the experimental FWHM of the $G^-$ peak of metallic SWNTs with the theroetical HWHM.

\bibitem{andoSWNTs}K.Ishikawa,T.Ando, J.Phys.Soc.Jap.\textbf{75}, 084713 (2006).

\bibitem{Note_FWHM2} Note that the prefactor of Eq.7 of
Ref.\cite{SimoneNaturematerials2007} should be $\frac{\omega_0\alpha'}{4c}$

\bibitem{note01}EPC($\Gamma$) is equivalent to $\langle G^2_{\bf
\Gamma}\rangle_F$ as defined in Ref.\cite{LazPRB2006}

\end{thebibliography}
\end{document}